# Virtual Inertia Scheduling for Power Systems with High Penetration of Inverter-based Resources

Buxin She, *Student Member, IEEE*, Fangxing (Fran) Li, *Fellow, IEEE*, Hantao Cui, *Senior Member, IEEE*, Jinning Wang, *Student Member*, *IEEE*, Qiwei Zhang, *Student Member, IEEE*, Rui Bo, *Senior Member*, *IEEE*

***Abstract*—This paper proposes a new concept called virtual inertia scheduling (VIS) to efficiently handle the high penetration of inverter-based resources (IBRs). VIS is an inertia management framework that targets security-constrained and economy-oriented inertia scheduling and generation dispatch of power systems with a large scale of renewable generations. Specifically, it schedules the proper power setting points and reserved capacities of both synchronous generators and IBRs, as well as the control modes and control parameters of IBRs to provide secure and cost-effective inertia support. First, a uniform system model is employed to quantify the frequency dynamics of the IBRs-penetrated power system after disturbances. Based on the model, the *s*-domain and time-domain analytical responses of IBRs with inertia support capability are derived. Then, VIS-based real-time economic dispatch (VIS-RTED) is formulated to minimize generation and reserve costs, with a full consideration of dynamic frequency constraints and derived inertia support reserve constraints. The virtual inertia and damping of IBRs are formulated as decision variables. To address the non-linearity of dynamic constraints, deep learning-assisted linearization is employed to solve the optimization problem. Finally, the proposed VIS-RTED is demonstrated on a modified IEEE 39-bus system. A full-order time-domain simulation is performed to verify the scheduling results**.

## I. Introduction

### A. Background

THE bulk power grid is transforming from a system dominated by synchronous generators (SGs) to one of hybrid SGs and inverter-based resources (IBRs). It's possible that the bulk power grid may become a 100% IBRs system in the future [1]. The transformation poses both opportunities and challenges to the power system [2]. On the one hand, IBRs make the best use of renewable energies, such as solar photovoltaics (PV), wind, tidal energy, and biomass energy, which are more environmentally friendly [3]. On the other hand, the integration of IBRs brings more uncertainty to the power grid and makes it difficult to predict and cooperate with the existing grid devices [4]-[5]. IBRs have lower physical inertia than conventional SGs and thus pose a threat to system frequency maintenance [6]. Hence, this manuscript focused on the development of an urgently needed inertia management framework for IBRs-penetrated power systems.

### B. Literature review

The existing literature that addresses low inertia issues can be roughly divided into two categories: i) methods that pursue inertia support from IBRs by designing new control strategies with inertia support capability; and ii) approaches that make better use of the inertia support capability of existing devices by integrating dynamic frequency constraints into an economic operation framework.

#### 1) IBR control algorithm with inertia support capability

Although IBRs have no rotating rotor mass, they can provide inertia support to the power system through elaborate control algorithm design. Therefore, the concept of synchronverters, or virtual synchronous generators (VSGs), is proposed to provide ancillary services of synthetic inertia support for electric power systems [7]. The grid-following IBRs track the frequency of the main grid and generate power reference using a proportional-differential controller, while the grid-forming IBRs measure the power out and generate frequency reference to avoid calculating differential terms that are sensitive to high frequency harmonics [8]. Fundamentally, VSG-controlled IBRs can provide virtual inertia support by increasing the active power injection rapidly right after a disturbance, under the guidance of pre-configured control strategies.

Apart from the basic VSG control method, some improved algorithms have been proposed to enhance the inertia support capability of VSG-controlled IBRs. For example, a $L_2$ and $L_{inf}$ norm controller was designed in [9] for VSG-controlled inverters, which improved IBRs' dynamic frequency response without increasing steady-state control effort. A linear-quadratic regulator based VSG was proposed in [10] for power systems with high inverter penetration, with the aim of making a tradeoff between the dynamic frequency constraints and the required control effort. In [11], an adaptive online virtual inertia and damping updating approach for virtual power plants was proposed to provide better inertia support for the main grid. In [12], a coordinated strategy for virtual inertial control and frequency damping control was proposed to explore the impact of virtual inertia and damping constant on frequency quality. An energy storage system, which has sufficient frequency reserves calculated by the final value theorem, was configured in [13] to provide inertia support for the main grid. In [14], a control strategy was proposed for the H-bridge converter with a combination of maximum power point tracking and VSGs. The selected reserved cells set aside a certain amount of the total PV power to act as a power buffer between the VSG and PV power, which helped keep the grid frequency stable.

Excluding VSGs, it has been determined that droop-controlled IBRs may have limited inertia support capability due to modules in the control loop, i.e., phase lock loop (PLL), low-pass filter, and so on [15]. In summary, the design of IBRs control methods with inertia support capability could relieve the problem of low inertia in future power systems.

*2) Frequency-constrained economic operation*

The conventional economic operation framework has three parts [16]: i). Unit commitment (UC) runs day-ahead scheduling to determine the unit's ON/OFF status; ii). Real-time economic dispatch (RTED) runs every 5 minutes to allocate the forecasted load among the committed units and schedule the reserve capacities; iii). Automatic generation control (AGC) is executed every 2-6 s to mitigate the frequency deviation and reduce tie line flow error. Then, security frequency constraints are integrated into the above framework to address the low inertia problem, considering the inertia support capability of grid devices. This contributes to the development of security-constrained UC, security-constrained RTED, and security-constrained AGC as follows.

*i). Security-constrained UC*: A frequency-constrained UC strategy that takes wind farm support into account was proposed in [17], using the uniform system frequency model [18] and piece-wise linearization. In [19], a stochastic UC strategy was developed for low-inertia grids, where the nonlinear dynamic constraints of RoCoF and nadir were transformed to bounded synthetic system parameters. In [20], a frequency reserve strategy was proposed for power systems under severe contingencies.

*ii). Security-constrained RTED*: Frequency constraints were considered in [21]. A confidence interval-based and distributionally robust RTED was proposed to strike a balance between security and economy. The operational risk was estimated based on wind power curtailment and load shedding resulting from wind power disturbance. A data-driven and distributionally robust optimization was developed in [22] for RTED, considering the secondary frequency regulation cost.

*iii). Security-constrained AGC*: An AGC constrained economic dispatch was formulated in [16] with full consideration of the short-term and long-term forecast. The adaptive and coordinated AGC strategy updated the regulation reserve online while guaranteeing the AGC variation constraints.

In general, the dynamic frequency response is gaining more and more attention in the economic operation of power systems.

## C. *Motivation and virtual inertia scheduling*

Following traditional practices, the device-level control algorithm design of IBRs and system-level economic operation have always been decoupled due to their distinct operational timescales. The former focuses on the dynamic control of power electronics to have the desired inertia support capability, while the latter usually works on demand-supply balance and cost-saving, assuming the preconfigured dynamic characteristics of the system devices. This is solid for conventional SG-dominated power systems because the response speed of SGs is relatively slow, and their parameters are usually fixed once configured. For example, the typical response time constant of the prime mover of a SG is 2 s and the start-up and shut-down time of a gas SG is 30 min. SGs' inertia constants and starting/closing time are determined by their physical configuration. However, the high penetration of IBRs challenges this assumption because IBRs have much faster electromagnetic responses than SGs and their control parameters can be changed adaptively in seconds or even milliseconds.

However, a double-edged sword cuts both ways. Although the high penetration of IBRs brings low inertia and quick dynamic features to power systems, it also opens the possibility of a more advanced scheduling framework by leveraging IBRs' controllability and flexibility. In existing frequency-constrained scheduling framework, IBRs are generally regarded as passive devices with constant control parameters. Motivated by better cooperation of device-level IBR control and grid-level economic operation, this paper proposes the concept of virtual inertia scheduling (VIS) for future low inertia power systems.

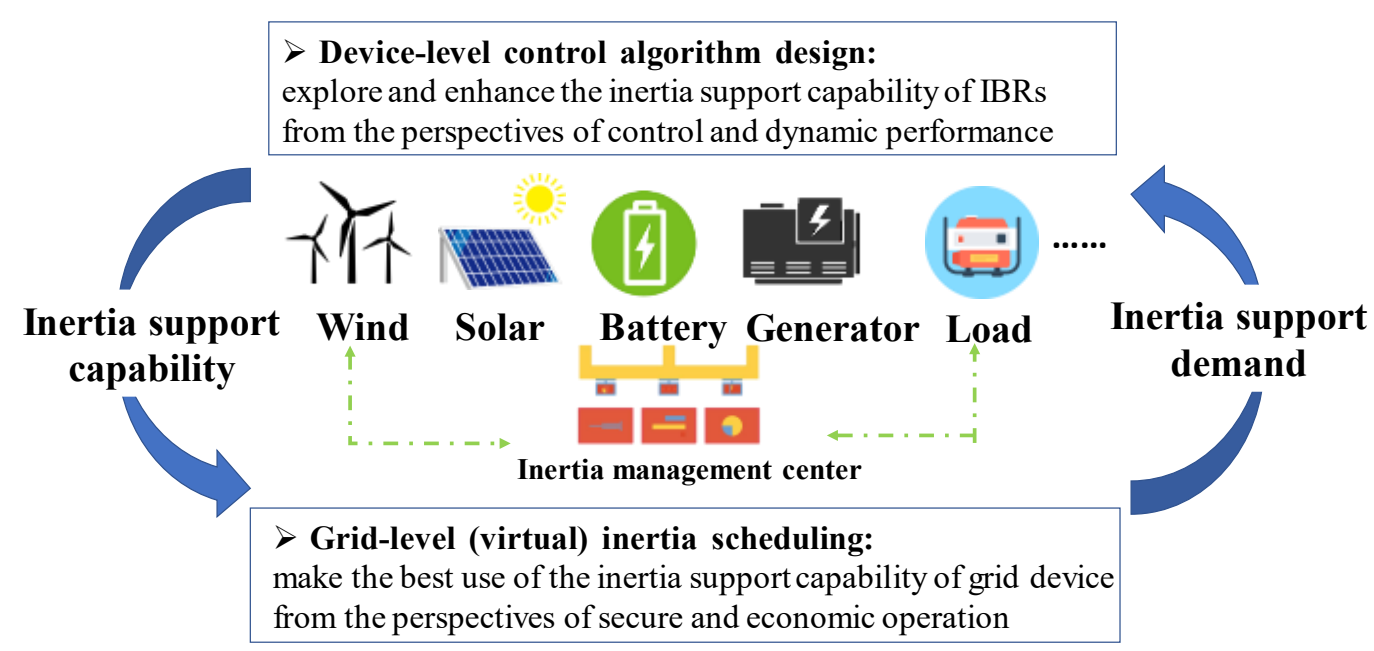

Fig. 1. Diagram of virtual inertia scheduling for future low inertia power systems.

VIS is an inertia management framework that targets security-constrained and economy-oriented inertia scheduling and power dispatch of power systems with a large scale of renewable generation. As shown in Fig. 1, the device-level control algorithm design explores the inertia support capability of IBRs and provides scheduling options (operation mode and operation parameters) for VIS. Then, VIS is set up at the grid-level to make the best use of the inertia support capability of IBRs. Compared with the conventional economic operation framework, VIS further addresses low inertia issues by leveraging the controllability and flexibility of power electronics-based devices that can respond quickly to scheduling results. It not only schedules the power setting points of system devices, but also determines the real-time operation modes and real-time control parameters of IBRs, as well as the required power reserve for inertia support.

VIS can fuse with each level of the economic operation framework, including UC, RTED, and AGC. To narrow down the research topic, this paper focuses on VIS-based real-time economic dispatch (VIS-RTED) for power systems with high penetration of IBRs. Specifically, we focus on a VIS-RTED that runs every 5 minutes to schedule the proper power setting points and reserve capacities of both SGs and IBRs, as well as the virtual inertia and damping of IBRs, to provide secure and cost-effective inertia support.

### *D. Contribution and paper organization*

In the following, we summarize three contributions of this paper:

- Proposes the concept of VIS, an inertia management framework that targets the security-constrained and economy-oriented inertia scheduling and power dispatch of power systems with a large scale of renewable generation.
- Derives the time-domain analytical response of IBRs with inertia support capability after disturbances, especially the analytical expressions of the power response, the peak power, and the time to reach the peak power.
- Formulates VIS-RTED for IBRs-penetrated low inertia power systems. The derived peak power is added as a power reserve constraint of IBRs, and the virtual inertia and damping are formulated as decision variables of the optimization problem. A deep learning assisted linearization approach is employed to deal with the nonlinear dynamic constraints.
- Conducts case studies to validate the formulated VIS-RTED. A full-order time-domain simulation is performed to obtain the dynamic response under the scheduling results, rather than using simplified transfer function models.

The rest of this paper is organized as follows. Section II introduces the dynamic frequency model of an IBR-penetrated power system. Section III derives the power dynamics of IBRs with inertia support capability, which is then integrated into the VIS formulation. The proposed VIS is integrated into RTED in Section IV, followed by a deep learning assisted approach to linearize the dynamic constraints. Then, Section V conducts case studies and makes comparisons with the existing methods. Conclusions are drawn in Section VI.

## II. Frequency Dynamics of IBR-penetrated Power System

This section introduces the modeling of IBR-penetrated power systems, including IBRs with inertia support capability, the uniform frequency dynamics model, and analytical frequency metrics.

### *A. Configuration of IBRs with inertia support capability*

Fig. 2(a) shows the configuration of a grid-following IBR connected to the main grid at the point of common coupling. Through elaborate controller design, IBRs can provide timely inertia support to the main grid, leveraging the quick response of power electronics. The controller consists of a primary regulator, a power regulator, and a current regulator. The primary VSG regulator emulates the physical characteristics of an SG by introducing a virtual inertia loop and a virtual damping loop. It degrades to a droop controller without considering the inertia emulation loop. Virtual inertia and damping determine the frequency support capability of the IBR, and they are two key decision variables in the VIS formulation.

With the help of a PLL, an IBR measures the frequency of the main grid, based on which the supplementary power signals are generated right after a disturbance. Considering the fast response of the PLL, current regulator, and power regulator, only the primary regulator is integrated into dynamics frequency modeling. The complete model in Fig. 2(a) will be used in the full-order time-domain simulation in Section V.

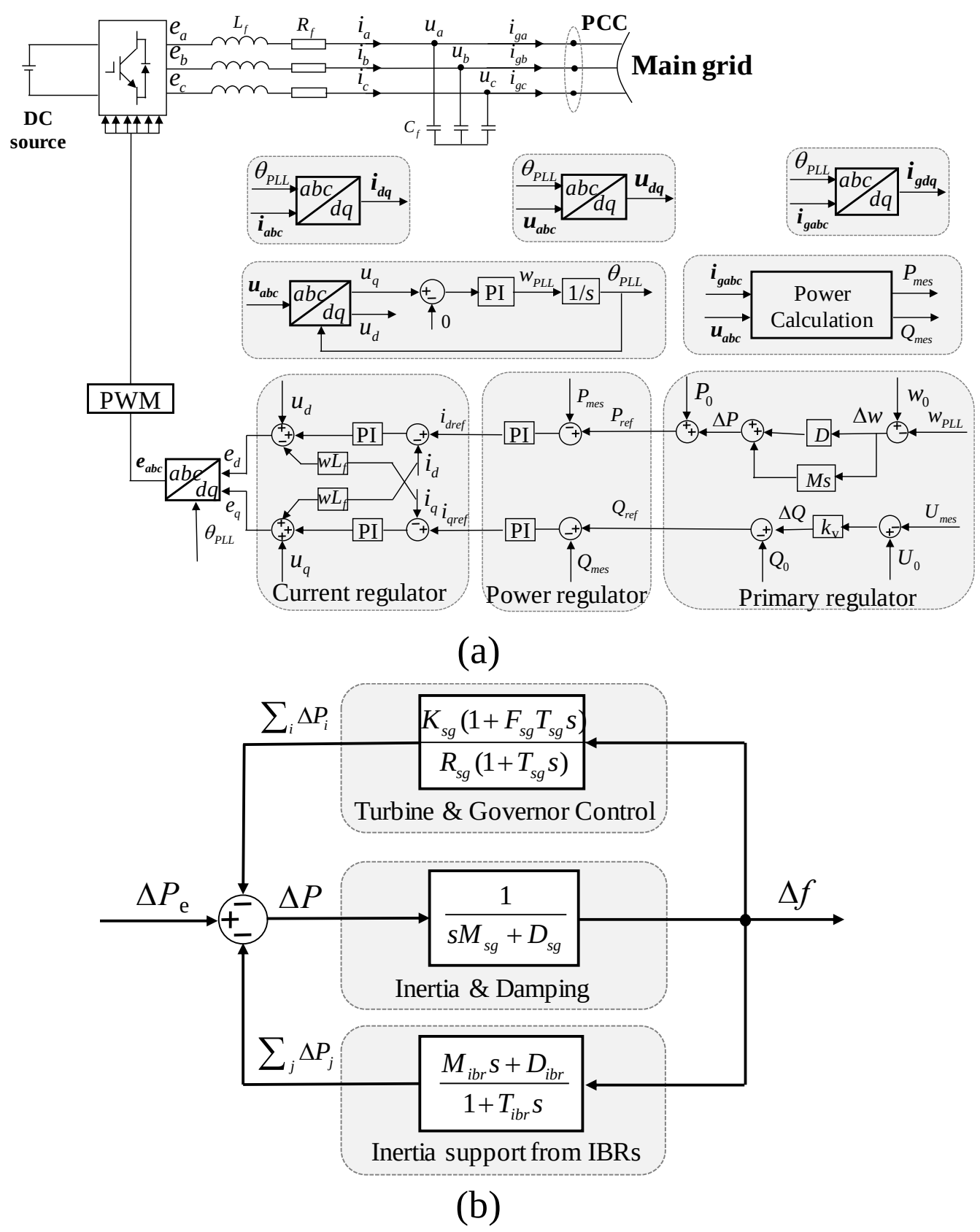

Fig. 2. Dynamic model of IBR-penetrated power system: (a) grid-following IBR with inertia support capability, and (b) uniform frequency dynamics model.

### *B. s-domain frequency response*

By referring to [10], this paper employs a simplified, but sufficiently accurate, uniform frequency dynamics model of an IBR-penetrated power system. As shown in Fig. 2(b), the uniform model considers the dynamics of the turbine, governor, and inertia support from IBRs. The closed-loop transfer function is shown in (1).

$$G(s)=\frac{\Delta f(s)}{\Delta P_e(s)}=\Bigg[\underbrace{(sM_{sg}+D_{sg})+\sum_{i=1}^{N_{sg}}\frac{K_{sg_i}(1+sF_{sg_i}T_{sg_i})}{R_{sg_i}(1+sT_{sg_i})}}_{\text{SGs}} +\underbrace{\sum_{j=1}^{N_{ibr}}\frac{sM_{ibr_j}+D_{ibr_j}}{1+sT_{ibr_j}}}_{\text{IBRs}}\Bigg]^{-1} \quad (1)$$

Assume all SGs have equal time constants ($T_{gi} = T$), and the inverter time constants are 2-3 orders of magnitude lower than $T$. Then, (1) is transformed as follows.

$$G(s) = \frac{\Delta f(s)}{\Delta P_e(s)} = \frac{1}{MT}\frac{1+sT}{s^2+2\zeta w_n s + w_n^2} \quad (2)$$

Where natural frequency $w_n$ and damping ratio $\zeta$ are calculated as follows. The synthetic parameters, i.e., inertia $M$, damping $D$, generation of fraction $F$, and droop $R$, can be found in [10].

$$w_n = \sqrt{\frac{D+R_g}{MT}}, \quad \zeta = \frac{M+T(D+F)}{2\sqrt{MT(D+R)}} \quad (3)$$

Assume a stepwise disturbance in the electrical power. Then the analytical $s$-domain frequency response is derived in (4).

$$\Delta f(s) = \frac{\Delta P_e}{MT}\frac{1+sT}{s(s^2+2\zeta w_n s + w_n^2)} \quad (4)$$

*C. Time-domain frequency response*

An inverse Laplas transform is performed in (4), then the analytical time-domain frequency response is derived as follows.

$$\Delta f(t) = \frac{\Delta P_e}{MTw_n^2}\left[1 - e^{-\zeta w_n t}\eta \sin(w_d t + \phi)\right] \quad (5)$$

where

$$w_d = \sqrt{1-\zeta^2}\, w_n, \quad \eta = \sqrt{\frac{1-2Tw_n\zeta + T^2 w_n^{\ 2}}{1-\zeta^2}}, \quad \tan\phi = \frac{w_d}{-Tw_n^2 + \zeta w_n} \quad (6)$$

The time instance of frequency nadir is determined by finding the instance at which the derivation of (5) is equal to 0. Then, the frequency nadir is derived by substituting the time instance into (5), and the maximum RoCoF occurs at t = $0^+$.

$$\Delta \dot{f}(t_m) = 0 \ \mapsto \ t_m = \frac{1}{w_b}\tan^{-1}\left(\frac{Tw_b}{\zeta T w_n - 1}\right) \quad (7)$$

$$\Delta f_{nadir} = \Delta f(t_m) = \frac{\Delta P_e}{MTw_n^2}\left[1 - \sqrt{1-\zeta^2}\,\eta e^{-\zeta w_n t_m}\right] \quad (8)$$

$$\dot{f}_{\max} = \Delta \dot{f}(0^+) = -\frac{\Delta P_e}{M} \quad (9)$$

## III. Power Dynamics of IBRs with Inertia Support Capability

IBRs require sufficient power reserves to provide secure inertia support. This section derives the analytical $s$-domain and time-domain power responses of VSG-controlled IBRs.

*A. Analytical power response*

The $s$-domain power response is obtained by integrating the $s$-domain frequency response into the feedback loop of VSG-controlled IBRs.

$$\Delta P_{ref}(s) = \frac{\Delta P_e}{MT}\frac{(M_{ivr}s + D_{ibr})(1+sT)}{s(s^2+2\zeta w_n s + w_n^2)} \quad (10)$$

By introducing some internal variables, (10) is simplified as

$$\Delta P_{ref}(s) = \frac{\Delta P_e D_{ibr}}{MTw_n^2}\left[-\frac{1}{s} + \frac{\alpha s + \beta}{s^2+2\zeta w_n s + w_n^2}\right] \quad (11)$$

where

$$\alpha = 1 - \frac{M_{ibr}Tw_n^2}{D_{ibr}}, \quad \beta = -Tw_n^2 + 2\zeta w_n - \frac{M_{ibr}w_n^2}{D_{ibr}} \quad (12)$$

Perform an inverse Laplas transform on (11) and combine the sine and cosine functions. Then, the time-domain power response is obtained as follows.

$$\Delta P_{ref}(t) = \frac{\Delta P_e D_{ibr}}{MTw_n^2}\left[-1 + \alpha\eta' e^{-\zeta w_n t}\sin(w_d t + \phi')\right] \quad (13)$$

where

$$\tan\phi' = \frac{w_d}{\beta/\alpha - \zeta w_n}, \quad \eta' = \sqrt{1 + \left(\frac{\beta/\alpha - \zeta w_n}{w_d}\right)^2} \quad (14)$$

Similar to the derivation of the frequency nadir, the time instance of peak power is determined by letting the derivation of (13) equal 0. The peak power is then calculated by plugging the time instance into (13).

$$\Delta \dot{P}_{ref}(t_m') = 0 \mapsto$$
$$t_m' = \frac{1}{w_d}\tan^{-1}\left(\frac{(\beta - 2\zeta\alpha w_n)w_d}{\zeta\beta w_n - \zeta^2 w_n^2\alpha + \alpha w_d^{\ 2}}\right) \quad (15)$$

$$\Delta P_{\max} = \Delta P_{ref}(t_m') = \frac{\Delta P_e D_{ibr}}{MTw_n^2}\left[-1 + \alpha\eta'\sqrt{1-\zeta^2}\,e^{-\zeta w_n t_m}\right] \quad (16)$$

*B. Power response curve*

Due to their controllability and flexibility, IBRs can provide inertia support to the main grid through elaborate controller design. Fig. 3 shows the typical time-domain response of frequency and power response based on (5), and (13). The observations are three-fold:

- Both the droop-controlled IBR ($M_{ibr}$=0) and the VSG-controlled IBR provide frequency support by increasing the active power injection to the main grid instantly after the step disturbance.
- The power response of a droop-controlled IBR has the same overall shape but the opposite trend to the frequency trajectory, and they reach the peak/nadir at the same time. This is because $\Delta P_e$ is calculated by timing frequency deviation and droop coefficients.
- Compared with droop-controlled IBR, VSG-controlled IBR has a much sharper power response instantly after the step disturbance but the same output at the steady state. Moreover, it has a larger peak power output and a shorter time to reach the peak. That's why the VSG-controlled IBR has much better inertia support capability.

SGs provide rotating inertia support to the main grid by releasing the energy stored in the rotors, which is more of a mechanical process with physical support. The inertia support of IBRs, on the other hand, is more of an electromagnetic process associated with the fast discharge of power electronics. Its inertia support energy is usually stored in the DC side capacitors or storage systems, and is a much smaller amount than the mechanical energy of the SG rotor. The insufficient power reserve of an IBR may easily result in a DC-side voltage dip and even generator trip [23]. Hence, it is important to set aside some inertia support reserves [14] or some headroom from the maximum power point [7] for IBRs that provide online inertia support, which motivates the derivation in this section.

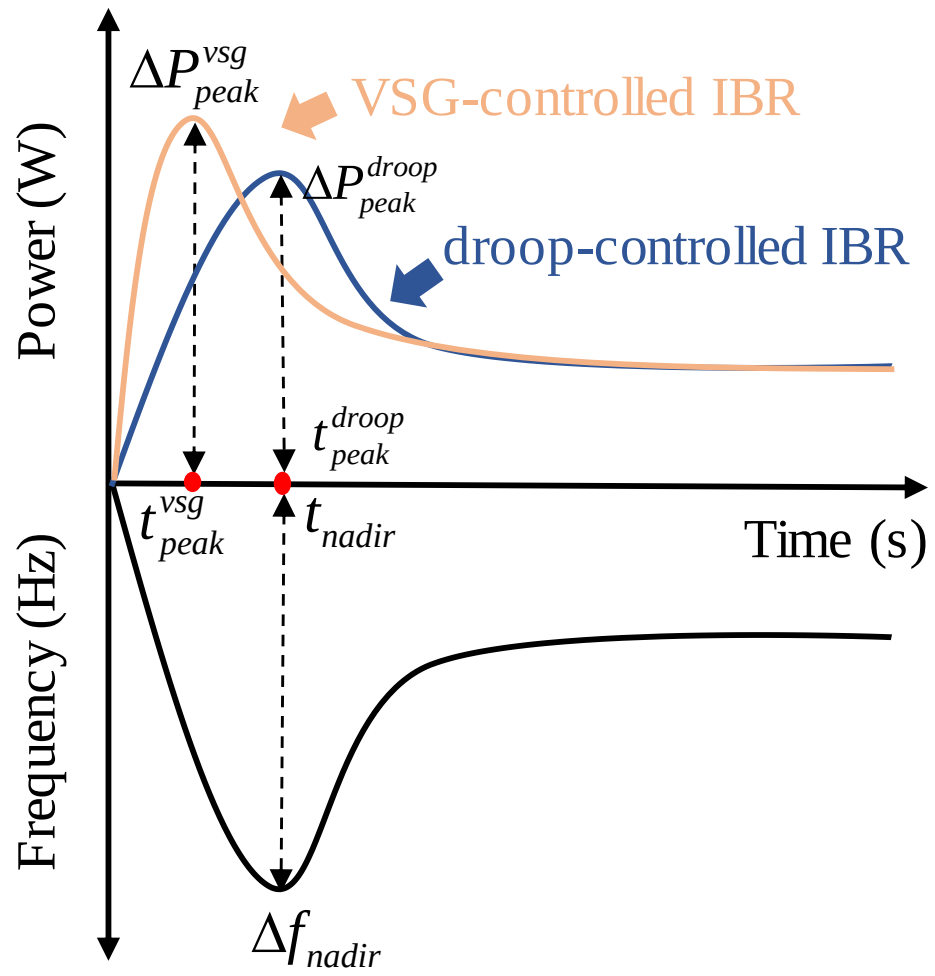

Fig. 3. Power and frequency responses after step disturbance.

## IV. Virtual Inertia Scheduling In Real-Time Economic Dispatch

This section integrates VIS into RTED. A deep learning assisted algorithm is then used to linearize the dynamic constraints in VIS-RTED.

### A. VIS based real-time economic dispatch

As shown in (17)-(23), VIS-RTED aims to minimize the total quadratic generation cost and linear reserve cost while also accounting for dynamic frequency constraints and IBR inertia support capability.

$$\begin{aligned}\min_{P,M,D}\sum_{t\in\mathbb{T}}[&\underbrace{\sum_{i=1}^{N_{sg}}(a^{sg}_{i,t}P^{sg\,2}_{i,t}+b^{sg}_{i,t}P^{sg}_{i,t}+c^{sg}_{i,t}+b^{sg}_{r,i,t}P^{sg}_{i,r,t})}_{SG}\\&+\underbrace{\sum_{i=1}^{N_{ibr}}(a^{ibr}_{i,t}P^{ibr\,2}_{i,t}+b^{ibr}_{i,t}P^{ibr}_{i,t}+c^{ibr}_{i,t}+b^{ibr}_{r,i,t}P^{ibr}_{i,r,t})}_{IBR}]\end{aligned} \tag{17}$$

$$s.t.\ \sum_{i=1}^{N_{sg}}P_{i,t}+\sum_{i=1}^{N_{ibr}}P_{i,t}+P_{i,t}-\sum_{i=1}^{N_l}L_{i,t}=0,\ \forall t\in\{1,\cdots,T\} \tag{18}$$

$$\begin{cases}\sum_{i=N}^{N_l}GSF_{l,t}(G_{i,t}+P_{s,i,t}-L_{i,t})\le LU_l\\ \sum_{i=N}^{N_l}GSF_{l,t}(G_{i,t}+P_{s,i,t}-L_{i,t})\ge -LU_l\end{cases} \tag{19}$$

$$\begin{cases}P^{sg}_{i,t}+P^{sg}_{i,ru,t}\le P^{\max,sg}_{i,t},\ \forall t\in\{1,\cdots,T\}\\ P^{sg}_{i,t}-P^{sg}_{i,ru,t}\ge P^{\min,sg}_{i,t},\ \forall t\in\{1,\cdots,T\}\end{cases} \tag{20}$$

$$\begin{cases}P^{ibr}_{s,i,t}+P^{ibr}_{i,ru,t}+P^{ibr}_{i,peak,t}\le P^{\max,ibr}_{i,t},\ \forall t\in\{1,\cdots,T\}\\ P^{ibr}_{s,i,t}-P^{ibr}_{i,rd,t}-P^{ibr}_{i,peak,t}\ge P^{\min,ibr}_{i,t},\ \forall t\in\{1,\cdots,T\}\end{cases} \tag{21}$$

$$\begin{cases}M^{\min,ibr}_i\le M^{ibr}_i\le M^{\max,ibr}_i,\forall i\in\{1,\cdots,N_{ibr}\}\\ D^{\min,ibr}_i\le D^{ibr}_i\le D^{\max,ibr}_i,\forall i\in\{1,\cdots,N_{ibr}\}\end{cases} \tag{22}$$

$$\begin{cases}-RoCof_{\lim}\le f_0\dfrac{\Delta P_{e,t}}{M_t}\le RoCof_{\lim},\ \forall t\in\{1,\cdots,T\}\\ f_{\min}\le f_0+\Delta f_{nadir,t}\le f_{\max},\ \forall t\in\{1,\cdots,T\}\end{cases} \tag{23}$$

where (18) is the generation and load balance equation; (19) is the transmission line thermal constraint; (20) is the SG generation constraint with regulation up or regulation down reserve; (21) is the IBR's generation constraint with up or down inertia support reserve; (22) is the virtual inertia and damping constraint of IBRs; (23) is the frequency nadir and RoCoF constraint.

Compared with conventional RTED, the proposed VIS-RTED has the following merits:

- VIS-RTED considers frequency nadir and RoCoF limits, which are critical for low inertia power systems with high penetration of IBRs.
- VIS-RTED formulates virtual inertia *M* and damping *D* as decision variables to make the best use of an IBRs' frequency regulation capability.
- VIS-RTED considers the inertia support reserve of IBRs, which reduces the risk of IBR DC voltage dip and generation trip.

In summary, VIS-RTED targets security-constrained and economy-oriented inertia management and real-time power dispatch, providing a good example of the feasibility of fusing VIS with the existing frequency regulated economic dispatch framework.

### B. Deep learning assisted linearization

Frequency nadir limits and inertia support reserves bring non-linear constraints to VIS-RTED, making it difficult for existing solvers to solve directly. As a result, a deep learning assisted linearization [24]-[25] approach is employed to linearize constraints (21) and (23). Two training datasets are generated within the feasible region of functions (5) and (13), with which two neural networks can be trained to predict the frequency nadir and IBR peak power. By introducing some binary variables and two large enough constants, the trained neural networks with activation function *ReLU* are transformed to mix-integer linear functions. Eqs. (24) and (25) show the expressions of the $m^{th}$ hidden layer before and after linearization, respectively.

$$\begin{cases}\hat{\boldsymbol{z}}_{\boldsymbol{m}}=\boldsymbol{W}_{\boldsymbol{m}}\boldsymbol{z}_{\boldsymbol{m-1}}+\boldsymbol{b}_{\boldsymbol{m}}\\ \boldsymbol{z}_{\boldsymbol{m}}=\max(\hat{\boldsymbol{z}}_{\boldsymbol{m}},\boldsymbol{0})\end{cases} \tag{24}$$

$$\begin{cases}\boldsymbol{z}_{\boldsymbol{m}}\le\hat{\boldsymbol{z}}_{\boldsymbol{m}}-\underline{\boldsymbol{h}}\odot(\boldsymbol{1}-\boldsymbol{a}_{\boldsymbol{m}})\\ \boldsymbol{z}_{\boldsymbol{m}}\ge\hat{\boldsymbol{z}}_{\boldsymbol{m}}\\ \boldsymbol{z}_{\boldsymbol{m}}\le\overline{\boldsymbol{h}}\odot\boldsymbol{a}_{\boldsymbol{m}}\\ \boldsymbol{z}_{\boldsymbol{m}}\ge\boldsymbol{0}\end{cases} \tag{25}$$

where $\odot$ means bitwise multiplication; $\boldsymbol{a}_{\boldsymbol{m}}$ is a binary vector; $\underline{\boldsymbol{h}}<\boldsymbol{0}$, $\overline{\boldsymbol{h}}>\boldsymbol{0}$, and [ $\underline{\boldsymbol{h}}$ , $\overline{\boldsymbol{h}}$ ] forms a vector interval that is large enough to contain all possible values of $\boldsymbol{z}_{\boldsymbol{m}}$.

Then, the two linearized networks can replace $P^{ibr}{}_{peak}$ and $\Delta f_{nadir}$ in (21) and (23). More transformation details can be found in [24] and [25]. Because the training datasets are generated based on the analytical expressions derived in Section III instead of case-by-case simulations, the trained neural networks have a high generalization capability and can be easily applied to other power systems while still retaining high prediction accuracy.

## V. CASE STUDY

This section conducts case studies to verify the formulated VIS-RTED. A full-order time domain simulation is performed to verify the scheduling results.

### *A. Case overview*

#### *1) Modified 39-bus system*

The test case is modified from the IEEE 39-bus system [26] with $S_{base}$=100MVA. As shown in Fig. 4, four SGs connected to Buses 30, 35, 37, and 38 are replaced by IBRs with capacities of 900 MW, 800 MW, 700 MW, and 1,000 MW, respectively. The case study assumes that the frequency nadir and RoCoF limits are 0.1Hz and 0.5 Hz/s [27], respectively. In addition, assume the maximum $M$ of an IBR is not larger than an SG with the same capacity, and the range of IBRs' virtual inertia and damping are at [0, 8.0] p.u. and [0, 6.0] p.u., respectively.

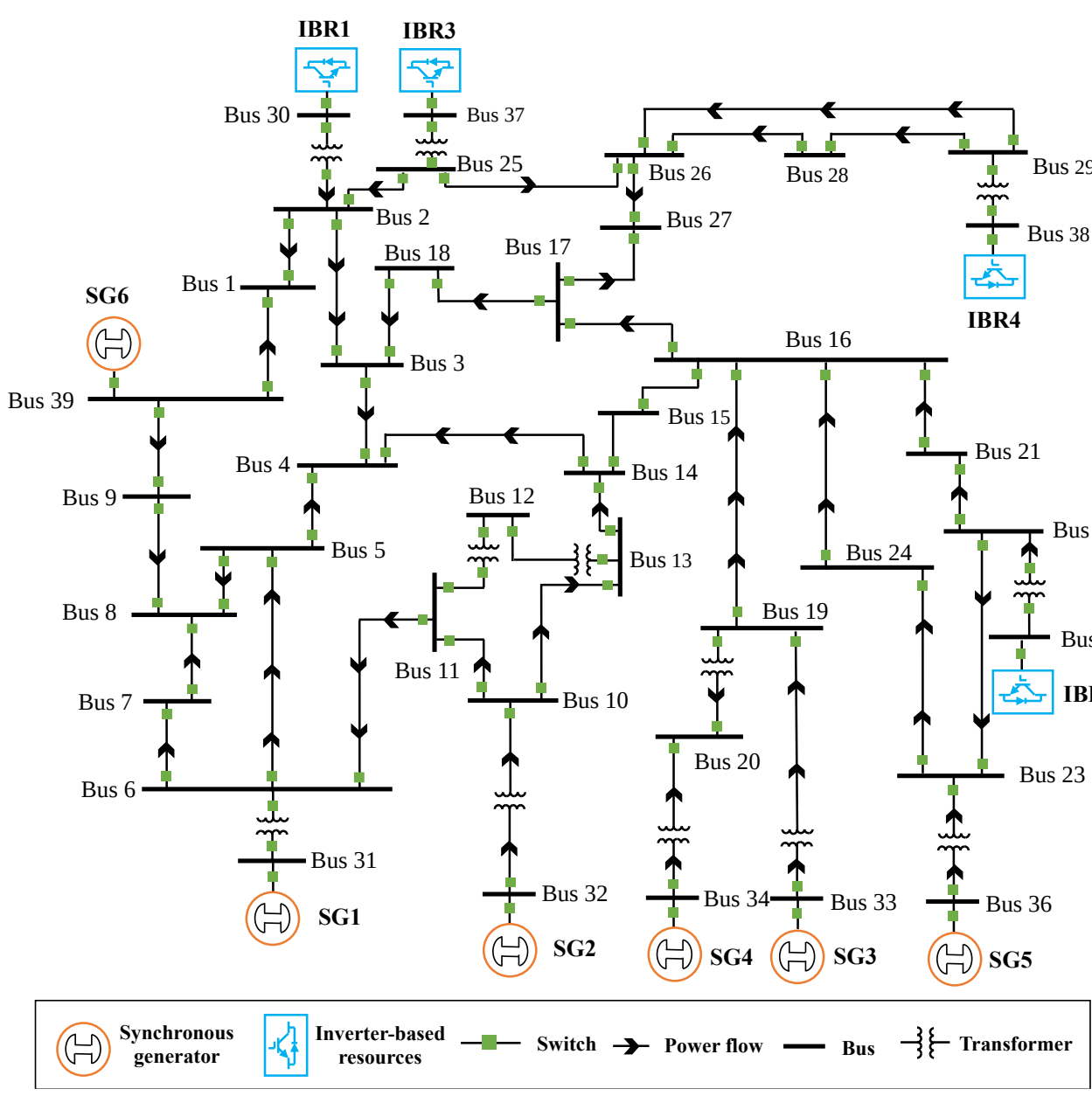

Fig. 4. One-line diagram of the modified 39-Bus system [26].

#### *2) Setup of VIS-RTED and time-domain simulation*

One-hour VIS-RTED ( $\mathbb{T}$ =**12**) will be solved on the modified 39-bus system. Assume that the SGs and IBRs have quartic fuel cost functions and linear reserve cost functions, respectively. The detailed cost data is shown in Table. I [28].

To emulate the operation of power systems as realistically as possible, VIS-RTED, AGC, and dynamic load change are integrated into time-domain simulation using *ANDES* [29]-[30], which is an open-source Python library for power system analysis and serves as the cornerstone for the *CURENT Large-scale Testbed (LTB)* [31]. The detailed dynamic simulation parameters can be found in the *ANDES* examples [32]. In each VIS-RTED interval, *ANDES* verifies the small signal stability of the scheduling results using the '*EIG*' function. *EIG* calculates the eigenvalues of the dynamic system. If the scheduling results produce a positive eigenvalue, IBRs' virtual inertia and damping will be fixed to their default values to guarantee stability.

Fig. 5. shows the one-hour load profile with a 1 s time-interval for time-domain simulation. The real-time load is synthesized as a forecasted load every 300 s, which is then fed to the VIS-RTED formulated in Section IV. As for the AGC model, the area control error that represents the system imbalance is regulated by a proportional-integral controller followed by a low-order power filter. Then, the area control signal is passed on to individual SGs every 4 s [16].

Table. I Cost data

| Generator ID | Generation cost | | | Reserve cost |
|---|---|---|---|---|
| | $a_g$($/MWh$^2$) | $b_g$($/MWh) | $c_{g,i}$($) | $b_r$($/MWh) |
| SG1 | 0.014 | 20 | 500 | 10 |
| SG2 | 0.020 | 20 | 380 | 10 |
| SG3 | 0.019 | 20 | 42 | 10 |
| SG4 | 0.020 | 20 | 380 | 10 |
| SG5 | 0.026 | 20 | 295 | 10 |
| SG6 | 0.021 | 20 | 400 | 10 |
| IBR1 | 0.001 | 1 | 50 | 20.61 |
| IBR2 | 0.001 | 1 | 50 | 18.96 |
| IBR3 | 0.001 | 1 | 50 | 19.15 |
| IBR3 | 0.001 | 1 | 50 | 20.06 |

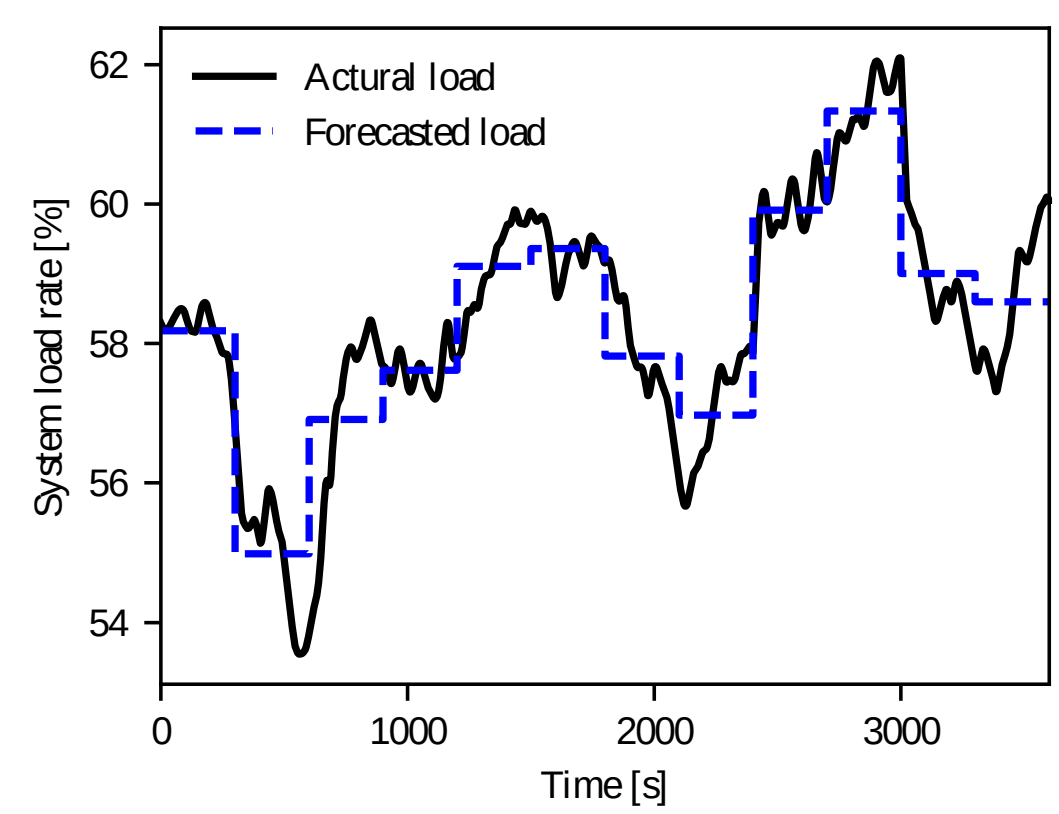

Fig. 5. One-hour Load profile.

### *B. VIS-RTED Results*

This subsection shows our deep learning training results, scheduling results, and time-domain simulation results.

#### *1) Deep learning training results*

Two multilayer perceptions are configured to predict the frequency nadir and VSG peak power, each with 1 hidden layer and 64 neurons. For each neural network, a dataset with a sample size of 20,000 was generated for training in PyTorch. Fig. 6 shows the training results, where (a) and (b) use a logarithmic scaled horizonal axis. After training for 1,000 epochs, the two networks can predict the frequency nadir and VSG peak power accurately, which means they can be integrated into the VIS formulation.

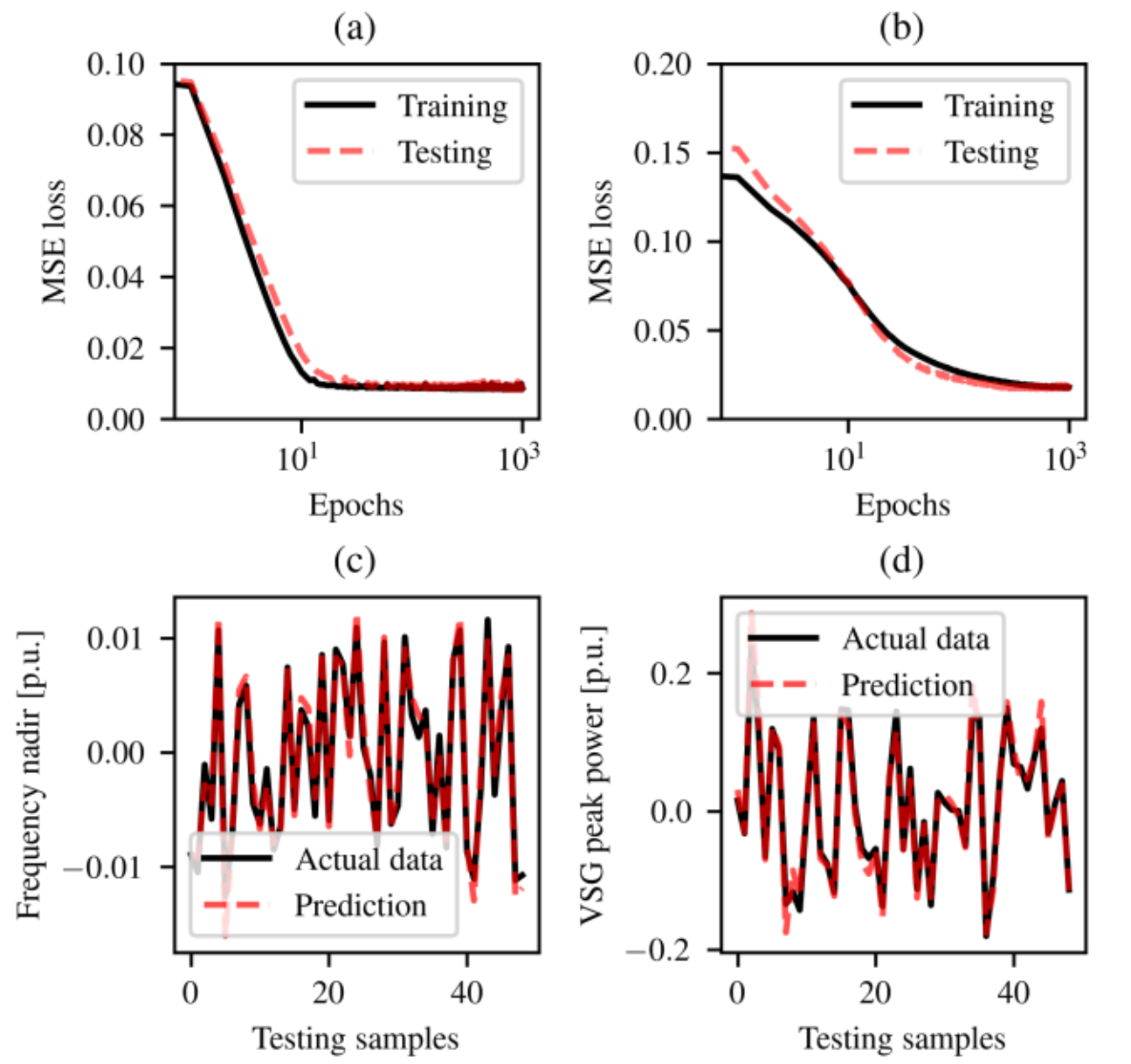

Fig. 6. DL training results: (a) Training loss of frequency nadir prediction; (b) Training loss of IBR peak power prediction; (c) Testing of frequency nadir prediction; (d) Testing of IBR peak power prediction.

*2) Scheduling results*

The total scheduling cost for a one-hour VIS-RTED is $63,300. Fig. 7 shows the detailed cost results of the 12 scheduling intervals, where (a) is the total system cost constituted by generation cost and inertia support reserve cost, and (b) is the cost of each SG and IBR in each scheduling interval. In Fig, 7(a), the inertia reserve cost is much smaller than the generation cost (fuel cost). This is because the inertia support reserve of a single interval is around 10% of disturbance $\Delta P_e$. In this paper, we just consider the normal load change, which is much smaller compared with total generation. If we consider large disturbance like generation trip, the cost of inertia support will increase significantly. In addition, because of the low cost of renewable energy, IBRs have a much lower generation cost than SGs in Fig. 7(b).

Fig. 8 shows the inertial support reserve alone with $\Delta P_e$. Referring to the dashed blue curve in Fig. 8(a), the 9$^{th}$ interval has the largest disturbance with $\Delta P_e$>0.04 p.u. Therefore, the 9$^{th}$ interval has the largest inertia support reserve, which is reflected in the total inertial reserve in Fig. 8(a) and the single IBR inertia support reserve in Fig. 8(b).

Fig 9 shows the virtual inertia and damping scheduling of each IBR, followed by the synthetic $M$ and $D$ of the whole system. Like in Fig. 8, the 9$^{th}$ interval has the largest virtual inertial scheduling result due to that interval containing the largest disturbance. In general, a higher $\Delta P_e$ necessitates a larger virtual inertia and thus a larger power reserve.

*3) Time-domain simulation results*

Fig. 10 shows the full order time-domain simulation results in *ANDES*. The observations are three-fold.

- During the one-hour time-domain simulation, the voltage and frequency in Figs. 10(a)-(b) and 10(e)-(f) are stable, demonstrating the stability of the VIS-RTED scheduling results, particularly the dynamic virtual inertia and damping of IBRs.
- RoCoF constraints are more critical compared with frequency nadir constraints under normal load change in low inertia power systems. The frequency curves are far from the up and down limits, as shown in Figs. 10 (b) and 10 (e), but the RoCoF of the IBRs reaches the limit around Fig. 10 (g).
- By comparing Fig. 10(c) and Fig. 10(g), the IBRs are shown to have larger RoCoF than SGs under the same disturbance.
- The scheduled output of IBRs is relatively stable and is higher than SGs due to the low cost of renewable energy resources.

In summary, the one-hour VIS-RTED is successfully solved and validated on the modified 39-bus system. With the proposed strategy, IBRs can provide secure and cost-effective virtual inertia support for the low-inertia power system, which contributes to the accommodation of more distributed energy resources.

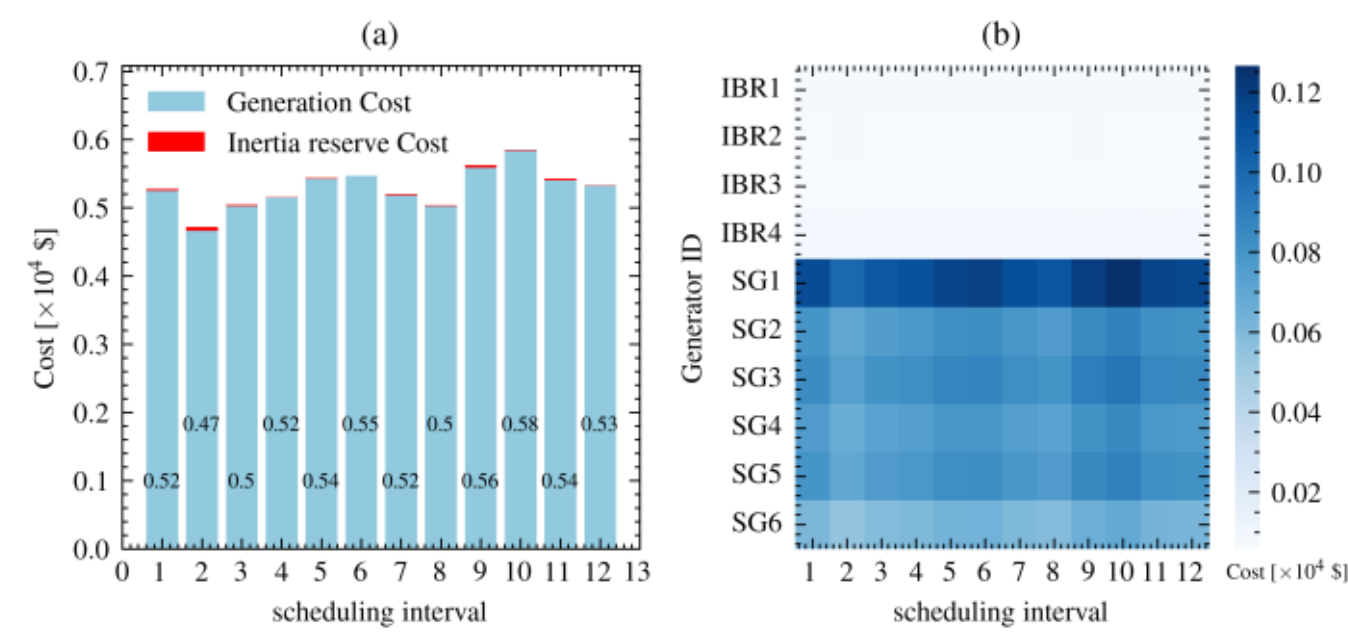

Fig. 7. VIS-RTED cost results: (a) system generation cost and inertia reserve cost; and (b) total cost of each generator.

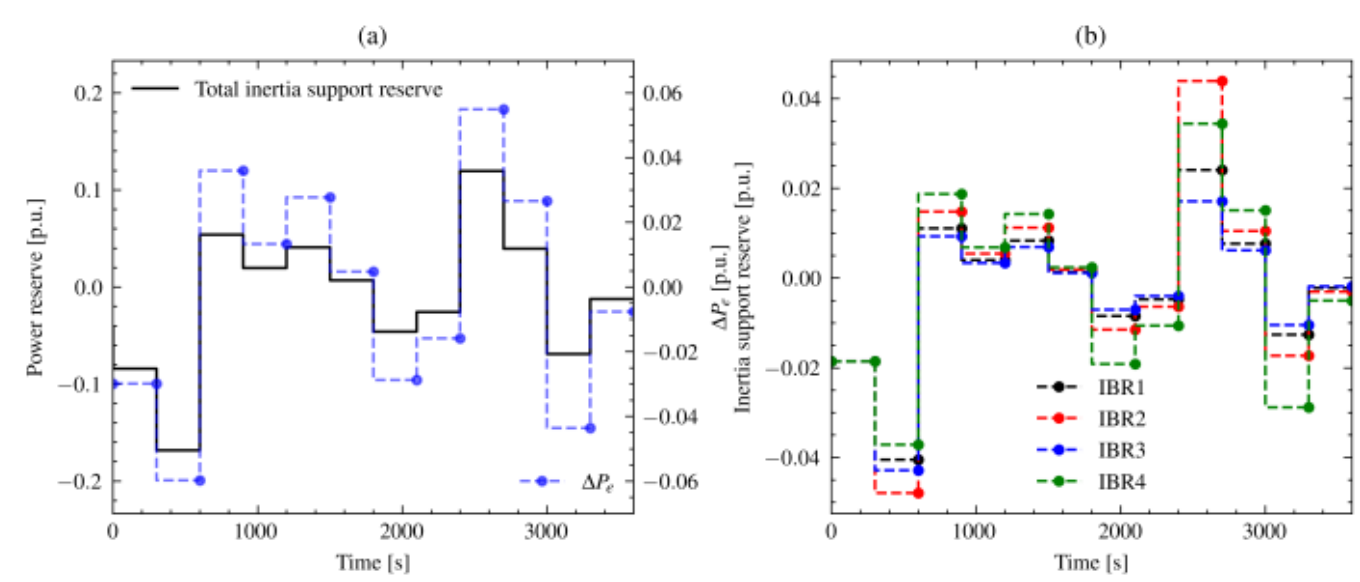

Fig. 8. IBRs inertia support reserve: (a) total reserve alone with $\Delta P_e$; (b) single IBR reserve.

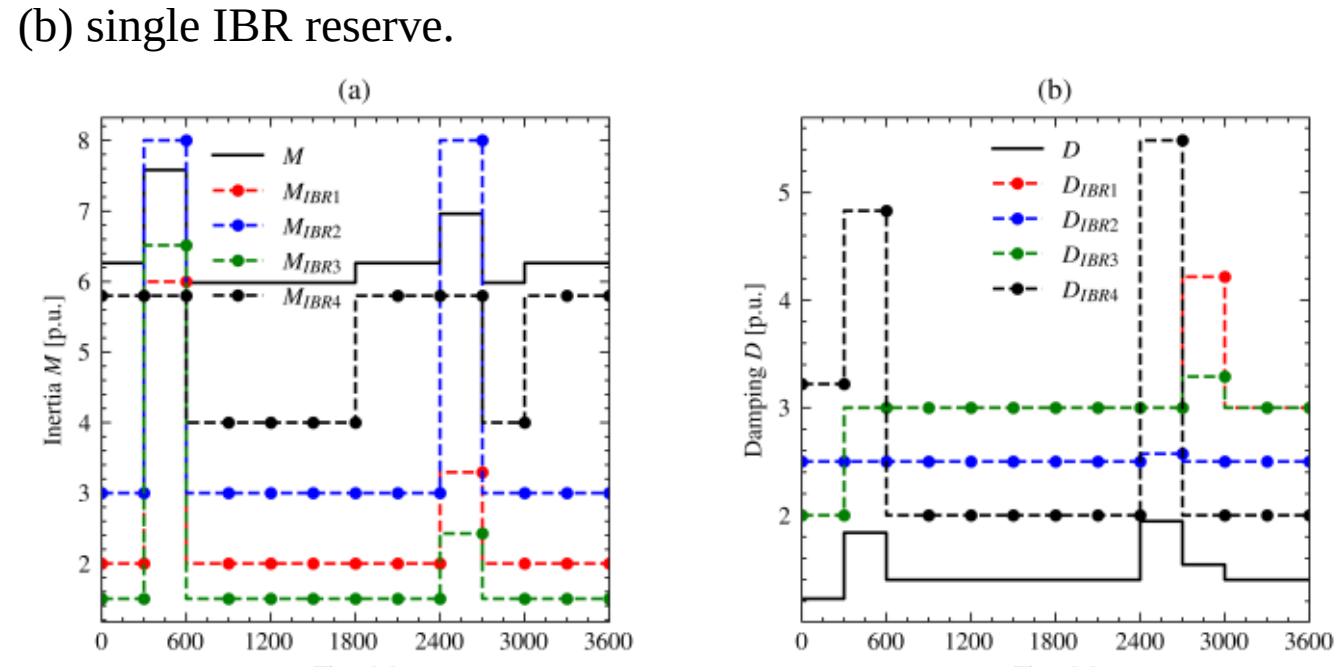

Fig. 9 IBRs inertia and damping: (a) virtual inertia; (b) virtual damping

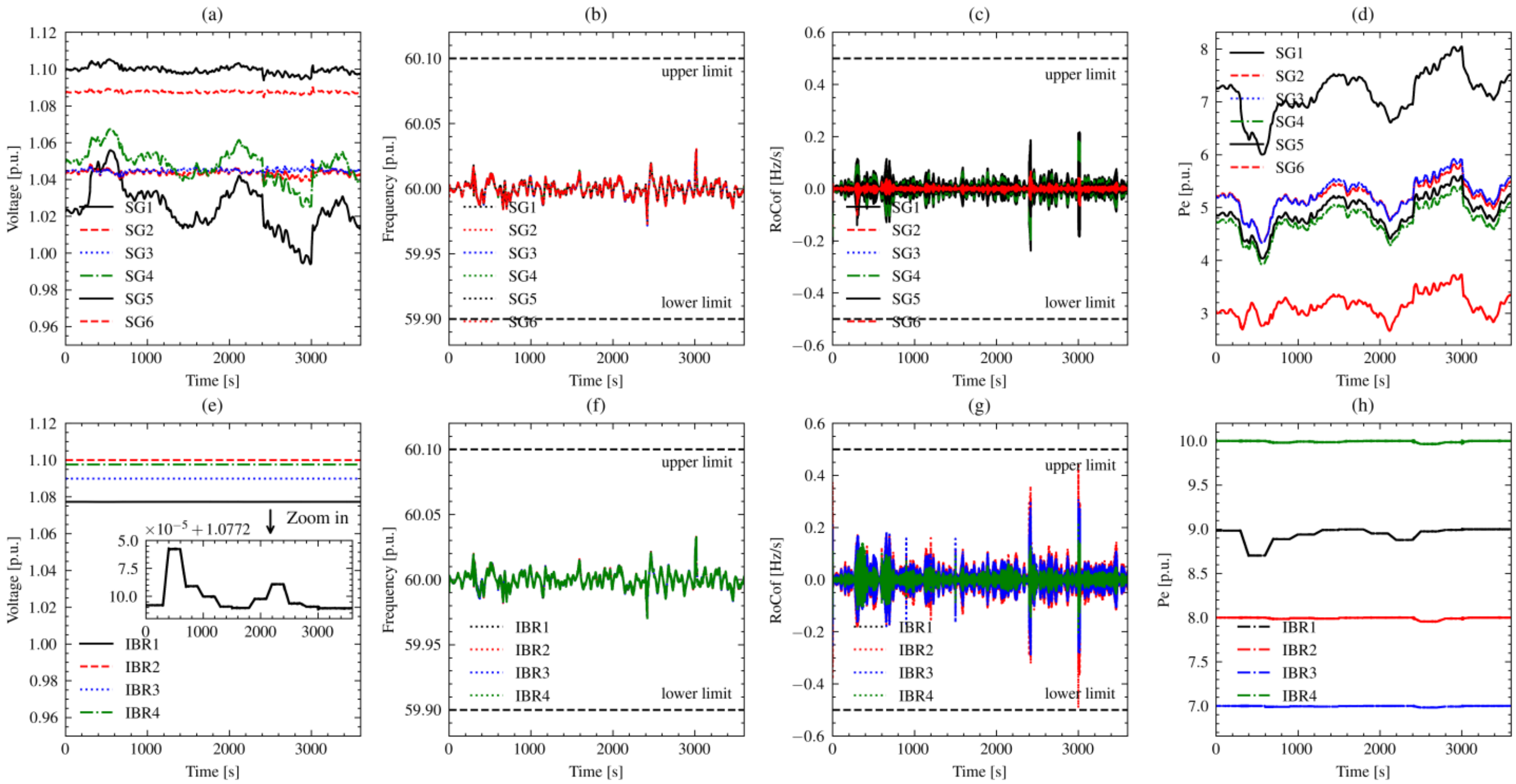

Fig. 10. Dynamics results through full-order time-domain simulation: (a) terminal voltage of SGs; (b) frequency of SGs; (c) RoCoF of SGs; (d) $P_e$ of SGs; (e) terminal voltage of IBRs; (f) frequency of IBRs; (g) RoCoF IBRs; and (h) $P_e$ of IBRs.

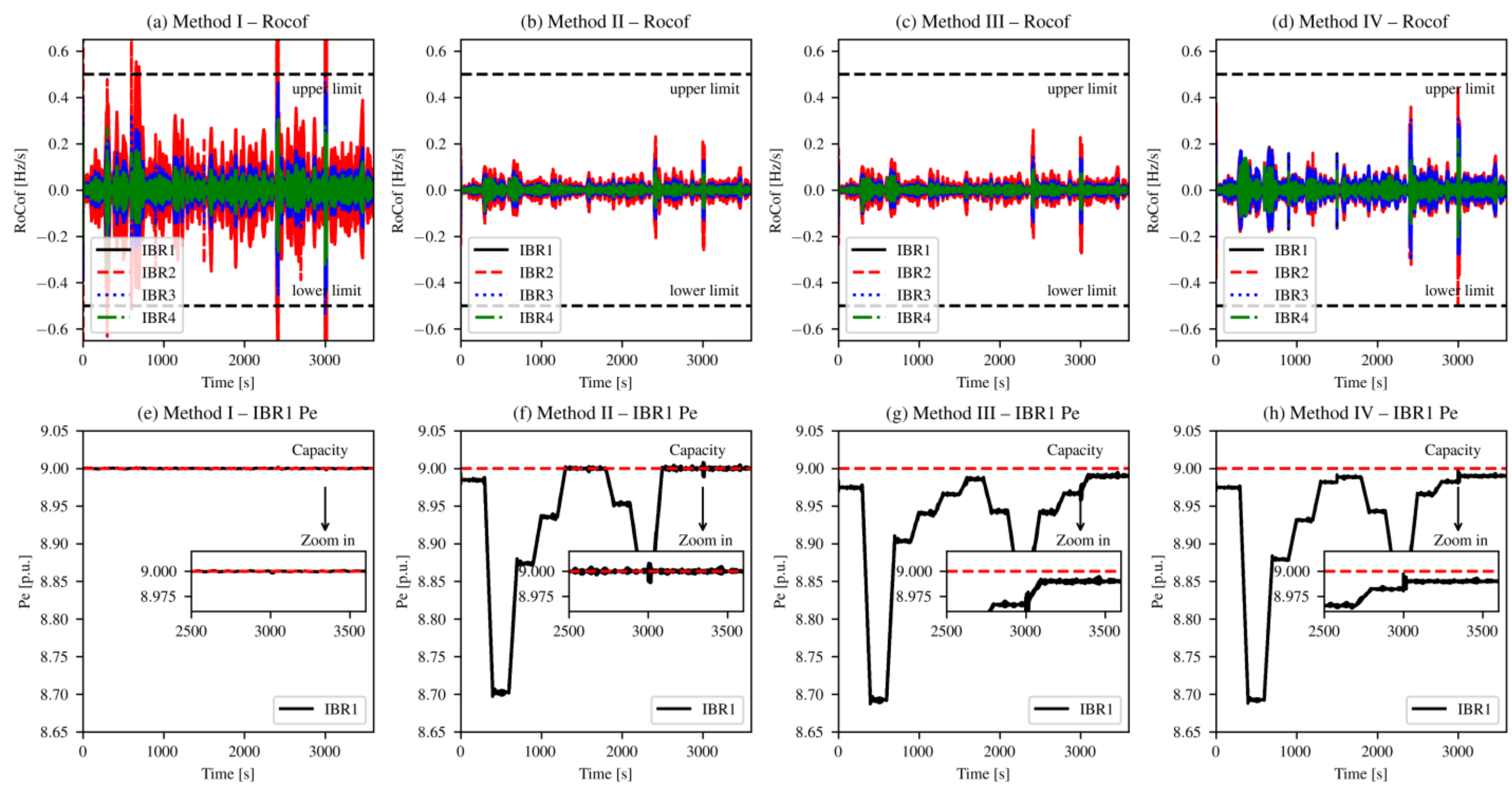

Fig. 11. Comparison of IBRs' RoCoF and IBR1's $P_e$ using different RTED methods: (a) Method I RoCoF; (b) Method II RoCoF; (c) Method III RoCoF; (d) Method IV RoCoF; (e) Method I IBR1 $P_e$; (f) Method II IBR1 $P_e$; (g) Method III IBR1 $P_e$; and (h) Method IV IBR1 $P_e$.

### C. *Performance analysis*

To better show the performance of the formulated VIS-RTED in Section IV, this paper chooses three baselines for comparison, using the same load profile as shown in Fig. 5. Table II shows the comparison results of the following methods. Two critical dynamic curves, i.e., all IBRs' RoCoF, and IBR1's $P_e$, are plotted in Fig. 11. The detailed setup of the four RTED methods is as follows.

- ***Method I***: ordinary RTED. IBRs work in in PQ control mode [33] with no inertia support capability.
- ***Method II***: ordinary RTED considering dynamic frequency constraints. IBRs work in VSG mode with fixed

virtual inertia and damping, but don’t have inertia support reserves.

- ***Method III***: VIS-RTED considering dynamic frequency constraints. IBRs have inertia support reserves, but with fixed virtual inertia and damping.
- ***Method IV***: complete VIS-RTED formulation in (17)-(23).

Table II Comparison of four RTED methods

| Index | I | II | III | IV |
|---|---|---|---|---|
| Total scheduling cost/×$10^4$\$ | 5.73 | 6.31 | 6.36 | 6.33 |
| Inertia support cost/×$10^4$\$ | 0 | 0 | 0.043 | 0.014 |
| Inertia support reserve/×$10^2$MW | 0 | 0 | 2.58 | 0.73 |
| Number of IBR capacity violations | 0 | **4** | 0 | 0 |
| Number of RoCoF violations | **4** | 0 | 0 | 0 |
| Number of frequency nadir violations | 0 | 0 | 0 | 0 |

As shown in Table II, the four RTED methods are compared from the perspective of scheduling cost and dynamic performance, based on which we have observed the following.

- Although *Method I* has the lowest total scheduling cost, it violates RoCoF limits 4 times. This is because PQ-controlled IBRs cannot provide any inertia support to the grid.
- *Method II* doesn’t violate any dynamic frequency constraints thanks to its security-constrained formulation. However, the output of IBRs temporarily exceeds the generation capacity 4 times during the transient process of inertia support.
- *Method III* solves the issue of IBR capability violation in inertia support when compared with *Method I* and *Method II*. However, it has higher inertia support costs and total scheduling costs than *Method IV* because the fixed $M_{ibr}$ and $D_{ibr}$ are determined based on the largest $\Delta P_e$, so as to guarantee the frequency performance under the worst scenario.
- Compared with *Method III*, the complete VIS-RTED with $M_{ibr}$ and $D_{ibr}$ as decision variables reduces the cost while maintaining sufficient inertia support power reserves. It outperforms the other three baseline methods.

The one-hour time-domain simulation results in Fig. 11 further verify the above observations. Focusing on the RoCoF curves in Fig. 11 (a)-(d), *Method I* has the worst results due to the absence of frequency support from IBRs; *Methods II-IV* have secure RoCoF curves, but *Method II* and *Method III* have smaller RoCoF curves than *Method IV* because IBRs have large fixed virtual inertia and damping. Focusing on IBR1’s $P_e$ curves in Fig. 11(e)-(f), *Method I* has constant IBR1 output with no inertia support capability; *Method II* sometimes breaks the capacity constraints of IBR1 because of the overshoot in the process of inertia support; and *Methods III* and *IV* strictly follow the capacity constraints of IBRs and have sufficient inertia support reserve.

In summary, the dynamic response of low inertia power systems is improved by leveraging the inertia support capability of IBRs. To provide sufficient and secure inertia support, it is necessary to set aside some IBR power reserves, which may introduce some extra costs. This calls for an advanced scheduling framework. The proposed VIS-RTED successfully integrates device-level IBR control parameter design into grid-level scheduling, resulting in an excellent tradeoff between economy and security. It also presents a good example of fusing VIS with the existing scheduling framework, which is beneficial to future low inertia power systems.

## VI. Conclusion

Although IBRs present low inertia characteristics, their controllability and flexibility allow for the design of an advanced inertia management framework for future low inertia power systems. Based on this background, this paper has proposed the concept of VIS, which targets the security-constrained and economy-oriented inertia management and power dispatch of power systems with large scale of renewable generation. VIS not only schedules the power dispatch results, but also the control modes and control parameters of system devices to provide secure and cost-effective inertia support.

The proposed VIS is integrated into RTED to perform online inertia scheduling every 5 minutes. VIS-RTED determines the power setting points and reserved capacities of both SGs and IBRs, as well as the virtual inertia and damping of IBRs, to provide sufficient and economic inertia support. Results show that VIS-RTED outperforms the existing RTED strategies in balancing cost-savings and security-enhancement. In the future, VIS will be added to the other frequency regulated economic dispatch frameworks, such as UC and AGC.

## VII. Acknowledgement

The authors would like to thank the financial support in part from the US DOD ESTCP program under the grant number EW20-5331 to complete this research work.